\documentclass{epl}
\newcommand{\bra}[1]{\langle #1|}
\newcommand{\ket}[1]{|#1\rangle}

\newcommand{\down}{\ket{{\downarrow}}}
\newcommand{\up}{\ket{{\uparrow}}}
\newcommand{\bradown}{\bra{{\downarrow}}}
\newcommand{\braup}{\bra{{\uparrow}}}

\usepackage{graphicx,amsmath,amssymb}
\usepackage{bm}

\title{Measurement of the ground-state flux
diagram of three coupled qubits as a first step towards the
demonstration of adiabatic quantum computation}
\shorttitle{Measurement of the ground-state flux diagram etc.}
\author{A. Izmalkov\inst{1}\thanks{E-mail: \email{andrei.izmalkov@ipht-jena.de}} \and M.~Grajcar\inst{2} \and S.~H.~W.~van~der~Ploeg\inst{1} \and U.~H\"{u}bner\inst{1} \and \\ E.~Il'ichev\inst{1}\thanks{E-mail: \email{ilichev@ipht-jena.de}} \and H.-G.~Meyer\inst{1} \and A.~M.~Zagoskin\inst{3,4}}
\institute{
  \inst{1} Institute for Physical High Technology, P.O. Box 100239, D-07702 Jena, Germany\\
  \inst{2} Department of Solid State Physics, Comenius University,\\ SK-84248 Bratislava, Slovakia\\
  \inst{3} Physics and Astronomy Dept., The University of British Columbia,\\
6224 Agricultural Rd., Vancouver, B.C., V6T 1Z1 Canada \\
   \inst{4} Frontier Research System, RIKEN, Wako-shi, Saitama, 351-0198,
   Japan }
\shortauthor{A. Izmalkov\etal} 

\begin{document}
\maketitle

\date{\today}

\begin{abstract}
The ground state susceptibility of a system consisting of three
flux-qubits was measured in the complete three dimensional flux
space around the common degeneracy point of the qubits. The
system's Hamiltonian could be completely reconstructed from
measurements made far away from the common degeneracy point. The
subsequent measurements made around this point show complete
agreement with the theoretical predictions which follow from this
Hamiltonian. The ground state anti-crossings of the system could
be read-out directly from these measurements. This allows one to
determine the ground-state flux diagram, which provides the
solution for the non-polynomial optimization  problem MAXCUT
encoded in the Hamiltonian of the three-flux-qubit system. Our
results show that adiabatic quantum computation can be
demonstrated with this system provided that the minimal energy gap
and/or the speed of the read-out is increased.

\end{abstract}

In the field of solid state qubits superconducting Josephson
junctions qubits are one of the most promising candidates for
quantum computation~\cite{Makhlin2001}. Currently they are
attracting considerable attention, mostly, because they are
potentially scalable, can be accessed relatively easily and
controlled individually. Quantum coherent oscillations and
conditional gate operation have been demonstrated in a two--qubit
system~\cite{PashkinYamamoto, McDermott2003}, entanglement between
the qubit and detector was achieved
recently~\cite{ChiorescuWallraff}, and a lot of attempts have been
made in the direction of improving the qubit's coherent dynamics
~\cite{Decoherence}. These experiments were aimed to construct a
universal set of gates as a basis for quantum computation.

An alternative approach is adiabatic quantum computation
(AQC)~\cite{Farhi2000}. It is based on the encoding of a
non-polynomially hard problem in a complex multi-qubit Hamiltonian.
This encoding should be done in such a way that the ground state
gives the solution of the problem. In order to reach this ground
state, adiabatic quantum evolution of the systems Hamiltonian is
used. As a preparation step, the system is moved to a flux
configuration where it can easily relax to the ground state. In the
subsequent calculation step, the system adiabatically evolves to the
configuration where its ground state encodes a non-polynomial hard
problem. In the final step, this ground state should be readout. A
scalable architecture for AQC using superconducting qubits was
proposed by Kaminsky \emph{et al.}~\cite{Kaminsky2004} and the
equivalence of AQC to standard quantum computation has been shown by
Aharonov \emph{et al.}~\cite{Aharonov2004}.

For flux qubits, such a controllable Hamiltonian can be built up by
making use of the flux dependence of their energies. Therefore, the
(energy) eigenvalues of this Hamiltonian become a N-dimensional
function of the flux through each qubit: $E_i(f_{q1}, f_{q2},...,
f_{qN})$ with $f_{qi}=\Phi_{qi}/\Phi_0-0.5$ being the flux
normalized by the flux quantum $\Phi_0=h/2e$ relative to the
degeneracy point of the qubit. The eigenstates of this Hamiltonian
can be described as a superposition of the $2^N$  states in the
natural qubit ($\down$ or $\up$) basis:
$\Psi=\alpha_1\mathopen|\downarrow_1\downarrow_2...\downarrow_N\rangle
+\alpha_2\mathopen|\uparrow_1\downarrow_2...\downarrow_N\rangle+...
+\alpha_{2^N}\mathopen|\uparrow_1\uparrow_2...\uparrow_N\rangle$. If
we set the fluxes far away from the degeneracy point of each qubit,
i.e. $\mathopen|f_{qi}\mathclose|>>0$, then the system is neither
entangled nor in a superposition of states but just in the simple
classical state:
$\alpha_j=1,~\alpha_1=...=\alpha_{j-1}=\alpha_{j+1}=...=\alpha_{2^N}=0$.
If the fluxes through all coupled qubits are close to their
degeneracies, then we approach the \emph{common} degeneracy point of
the whole system, where its state is maximally entangled. Close to
this degeneracy point the N-dimensional energy surface has a rich
structure with significant curvature~\footnote{Usually in the common
degeneracy point the spacing between ground $E_0$ and first excited
state $E_1$ (energy gap) is small in comparison with other flux
subspaces. The value of this gap is proportional to the qubits'
tunnelling amplitudes $\Delta_i$ and inversely proportional to the
interaction strengths $J_i$ between them. As the excitation
probability is inversely proportional to the energy gap the system
can get excited at this point and AQC will give a wrong answer.
Therefore, one should increase the energy gap by optimizing the
sample parameters. Also the evolution speed should be increased in
order to avoid thermal excitations while still being slow enough to
prevent Landau--Zener transitions.}. From this curvature, which is
directly proportional to the system's susceptibility, the qubits'
state can be determined~\cite{Grajcar2005}. In this Letter we show
the complete characterization of the susceptibility of a three flux
qubit system around its common degeneracy point.

The susceptibility can be determined by using a radio-frequency tank
circuit. In this technique the tank circuit plays the role of a
parametric transducer, since any variation of the real part
susceptibility $\chi$ of, for instance, the superconducting
interferometers~\cite{Silver67, Ilichev02} is transformed into a
change of the tank's resonance frequency. The variation of the
resonance frequency can be detected by measuring the phase shift
$\theta$ between the alternating current $I_{\mathrm{rf}}$ applied
through the coil and alternating voltage over the tank circuit.
These quantities are related through
\begin{equation} \tan \theta= -
\frac{Q_\mathrm{T}}{L_\mathrm{T}} \chi,
\label{formula1}\end{equation} where $L_\mathrm{T}$ is the tank
coil inductance and $Q_\mathrm{T}$ its quality.

For an \textit{N}-qubit system in the ground state $\chi=2
\sum_\nu\frac{r_{0\nu}}{ E_\nu - E_0 }$, where  $r_{0\nu}=\\
\left(\sum_{i=1}^N\lambda_i \langle 0|
\sigma_{z}^{(i)}|\nu\rangle\right) \nonumber \times
\left(\sum_{j=1}^N\lambda_j\langle
\nu|\sigma_{z}^{(j)}|0\rangle\right)$,
$\lambda_i=M_{qi\mathrm{T}}I_{\mathrm{p}i}$, $M_{qi\mathrm{T}}$ is
the mutual inductance between qubit \textit{i} and the tank,
$I_{\mathrm{p}i}$ is the persistent current of qubit
\textit{i}~\cite{Grajcar2005}. The susceptibility $\chi$ is
significant near the anti-crossings of the energy levels. These
anti-crossings are located close to the degeneracies of the
classical states ($\Delta_i=0$), e.g. $E(\uparrow
\uparrow\downarrow)=E(\downarrow\uparrow\downarrow)$ for three
qubits. Therefore this method allows one to find the location of
classical crossings in the full flux space of a multi qubit system.
For finite temperatures, in addition, the higher states become
populated thereby allowing the readout of their
anti-crossings~\cite{Grajcar2005a}.

The effective Hamiltonian for three coupled qubits is
\begin{equation}
 H= -\sum_{i=1}^3 [\epsilon_i \sigma_z^{(i)} + \Delta_i
 \sigma_x^{(i)}] + \sum_{1\le i<j\le3} J_{ij}
 \sigma_z^{(i)}\sigma_z^{(j)}\;,\label{H}
 \end{equation}
with $\epsilon_i$ the energy  bias on qubit~$i$, $\Delta_i$ its
tunnelling amplitude, $J_{ij}$ the coupling energy between qubits
$i$ and~$j$, and $\sigma_z$, $\sigma_x$ are the Pauli matrices in
the natural basis of $\down$ ($\bradown\sigma_z\down=-1$) and $\up$
($\braup\sigma_z\up=1$). Depending on the sign of $J_{ij}$ such a
Hamiltonian can describe systems with
anti-ferromagnetic\cite{Izmalkov2004a,Majer2005} and
ferro-magnetic\cite{You2005} interactions as well as a system with
both types of interaction~\cite{Grajcar2005b}. The sample described
here has only anti-ferromagnetic interactions. As a result the
system becomes frustrated around its common degeneracy point. In
fact, in the classical limit ($\Delta_i\rightarrow 0$) the
Hamiltonian~(\ref{H}) encodes the non-polynomial MAXCUT
problem~\cite{Grajcar2005, Steffen2003}. The solution of this
optimization problem is the classical state with the lowest energy.
By using the adiabatic evolution of the quantum system (\ref{H})
with $\Delta_i \neq 0$ one can find all anti-crossings of the ground
state and reconstruct the classical cross-overs from them. As was
explained in Ref.~\cite{Grajcar2005}, this information allows one to
determine the classical state which corresponds to the solution of
MAXCUT problem \footnote{ as an example, for the present system with
$f_{q1}=0.008,~ f_{q2}=0.02$ and $f_{q3}=0$ the solution of MAXCUT
problem is described by the state $|\uparrow \uparrow\downarrow
\rangle$, see fig.~\ref{fig3}b.}.

\begin{figure}[t!] \centering
\includegraphics[width=4cm]{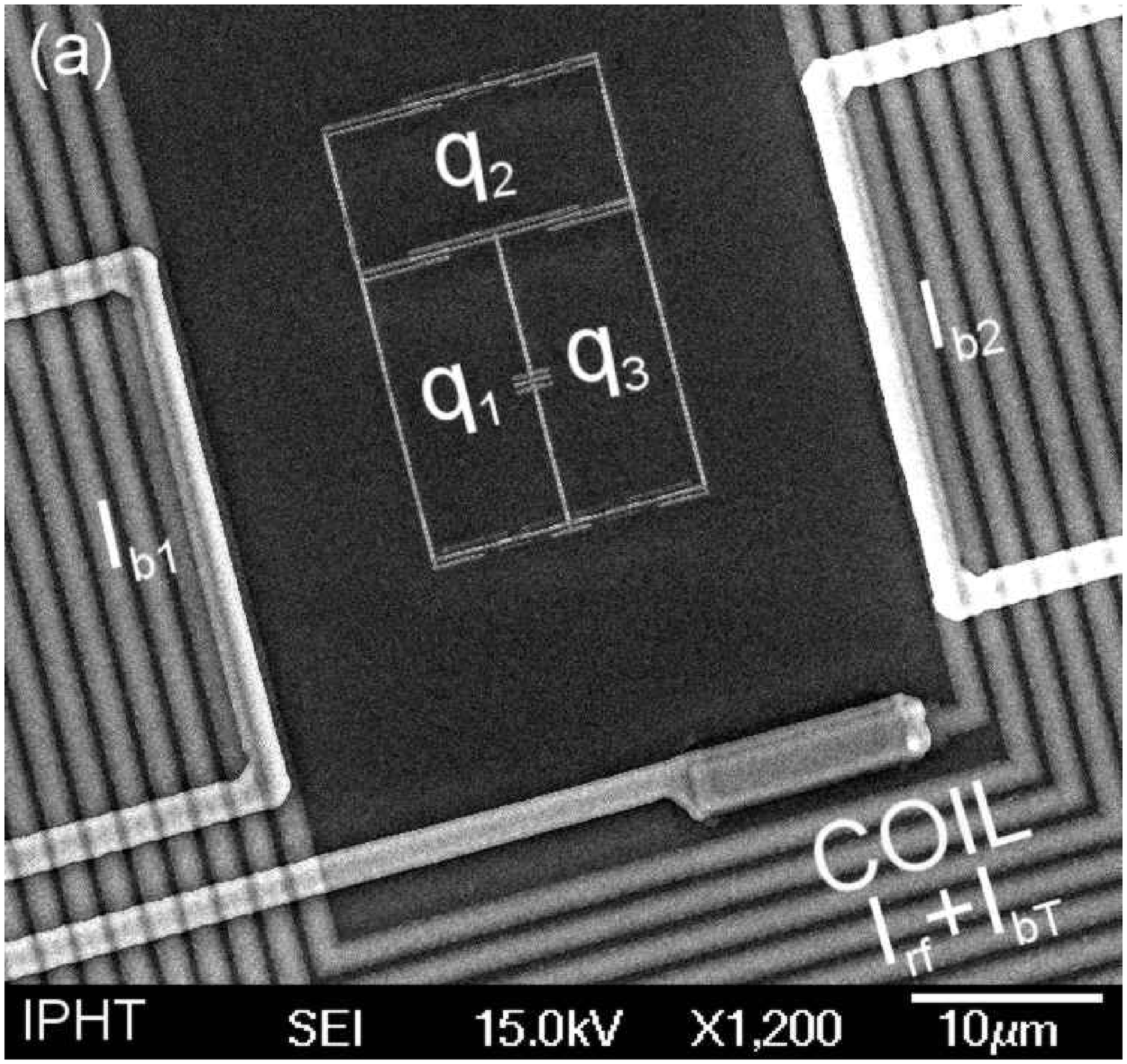}  \centering
\includegraphics[width=3.1cm]{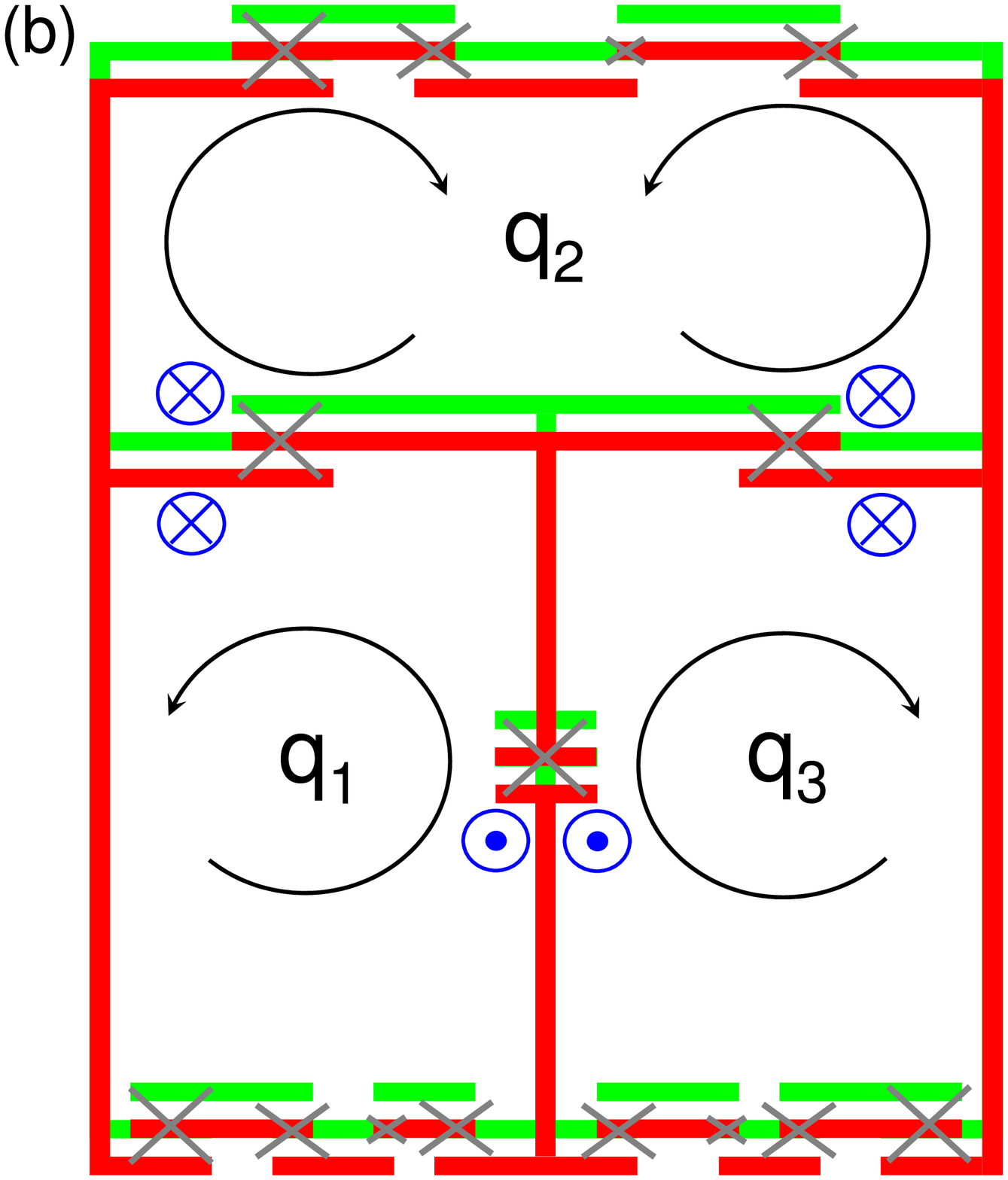} \caption{(Colour on-line) (a) Electron
micrograph of the sample, (b) layout of three coupled qubits. Each
$14\times7~\mu$m$^2$ loop contains six Josephson junctions. Two of
them ($120 \times 2000$~nm$^2$) are shared and couple the qubit to
its neighbors, three junctions ($120 \times 600$~nm$^2$, $120 \times
390$~nm$^2$, $120 \times 600$~nm$^2$) form the qubit and one
auxiliary junction ($120 \times 2000$~nm$^2$) is used to avoid
parasitic junctions. Green (light gray) and red (dark gray) layers
are bottom and top Al layers respectively. The arrows indicate the
directions of persistent currents in each qubit loop when they flow
trough the coupling junctions as indicated by the $\bigodot,
\bigotimes$ symbols. Since it is energetically preferable to have
the same direction of persistent currents through the coupling
junction the qubits become coupled in this case. So one can check
that all qubits pairs are anti-ferromagnetically coupled in this
design.} \label{fig1}
\end{figure}

Figure~\ref{fig1} shows the sample measured which consists of three
Al persistent current qubits~\cite{MooijOrlando1999} placed inside a
Nb pancake coil. Two junctions in each qubit are nominally
$600\times120$~nm$^2$, while a third one is $\sim$35\% smaller. Each
qubit is coupled to the other two both magnetically and via shared
$120 \times 2000$~nm$^2$ junctions~\cite{Levitov2001, Grajcar2005a}.
In order to prevent the formation of junctions in the leads, which
would lead to parasitic coupling, we add one auxiliary large
junction in each qubit circuit. The addition of three large
junctions increases the effective inductance of the loop and
modifies the Josephson potential of the qubits. In order to
compensate for this effect the size of the smallest junction has
been reduced. In comparison with previous designs with only one
coupling junction~\cite{Grajcar2005,Grajcar2005a} this allows to
couple all pairs anti-ferromagnetically.(see caption of
fig.~\ref{fig1} for details). The flux through the qubits can be
individually controlled by direct currents through the coil
$I_{\mathrm{bT}}$, and two additional lines $I_{\mathrm{b1}}$ and
$I_{\mathrm{b2}}$. The Nb coil has an inductance
$L_\mathrm{T}=134$~nH, and together with an external capacitance
$C_\mathrm{T}=470$~pF forms a parallel tank circuit with
$\omega_\mathrm{T}/2\pi=20.038$~MHz and quality
$Q_\mathrm{T}=\omega_\mathrm{T} R_\mathrm{T} C_\mathrm{T}=700$ (here
$R_\mathrm{T}$ is the effective resistance). The qubits were
fabricated by $e$\nobreakdash-beam lithography and two-angle shadow
evaporation, whereas the Nb coil was made by $e$-beam lithography
and CF$_4$ reactive-ion etching.

\begin{figure}[tbp]
\centering \includegraphics[width=5.5cm]{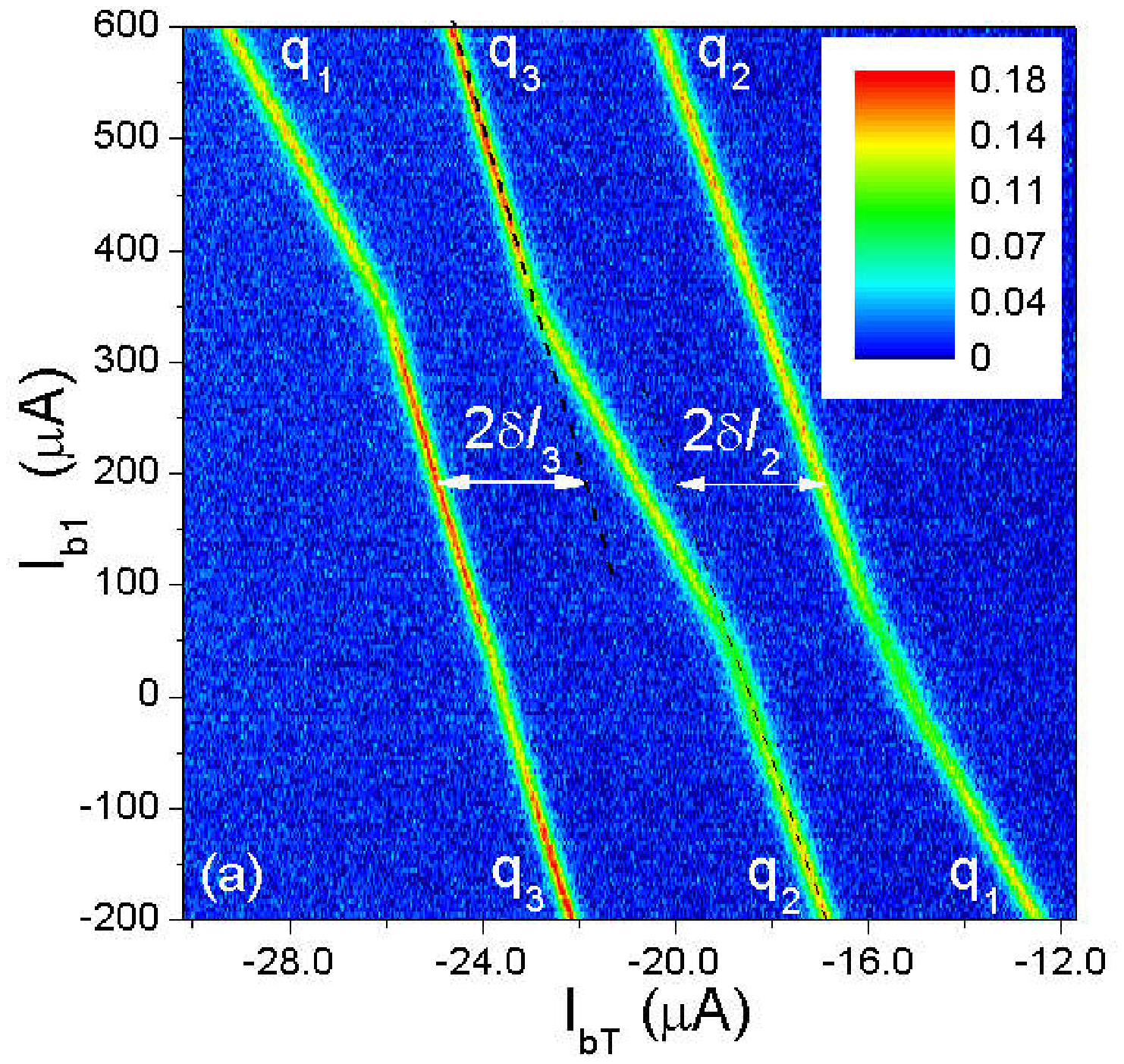} 
\includegraphics[width=5.5cm]{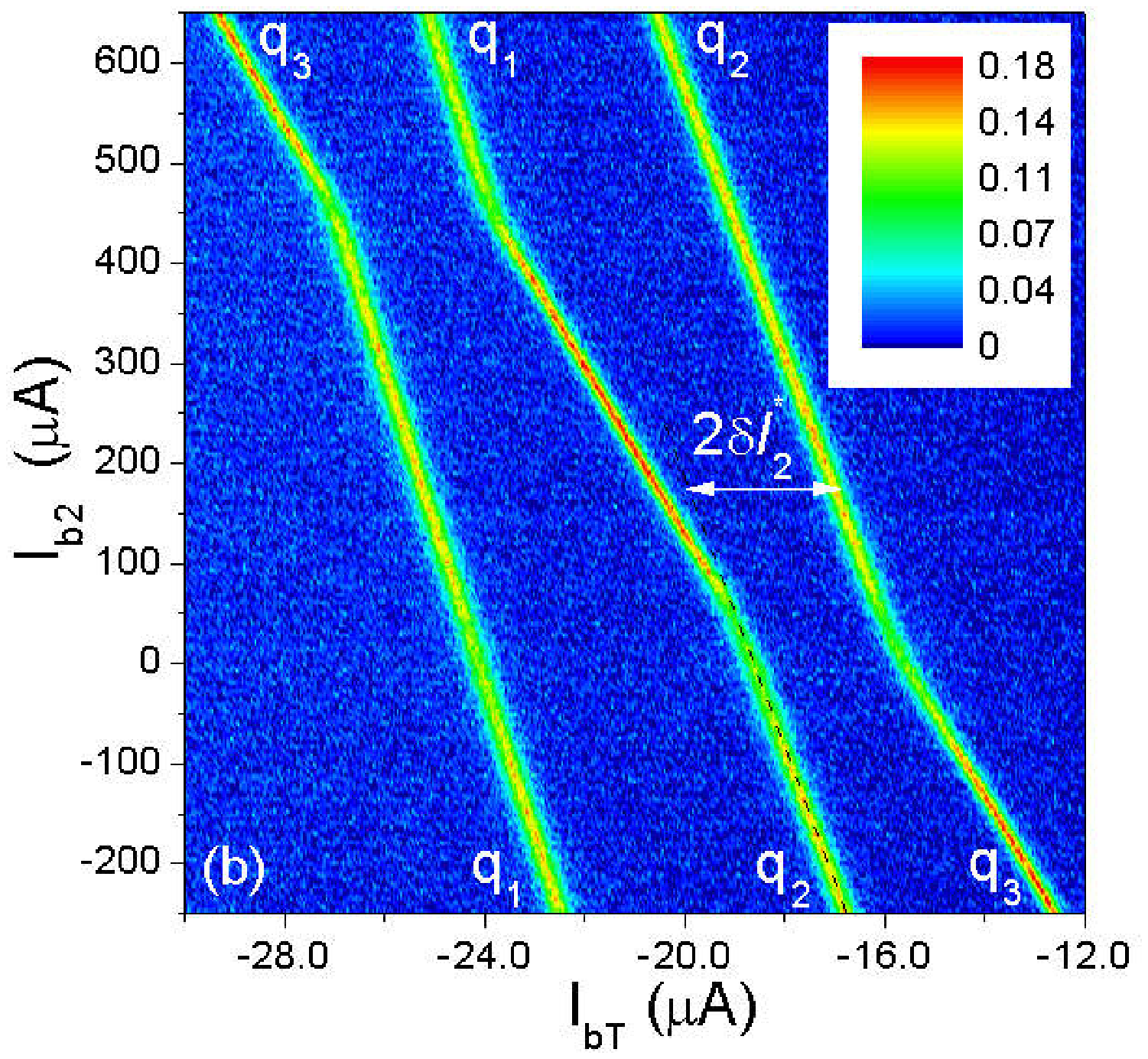} \caption{(Colour on-line) (a)
$-\tan \theta (I_{\mathrm{bT}}, I_{\mathrm{b1}})$ at
$I_{\mathrm{b2}}=400$~$\mu$A. The qubits' coupling strength
$J_{12}/I_{p2}\Phi_0 = M_{q2\mathrm{T}} \delta I_2/\Phi_0 \approx
0.034$ and $J_{13}/I_{p3}\Phi_0 = M_{q3\mathrm{T}} \delta I_3/\Phi_0
\approx 0.033$. (b) $-\tan \theta (I_{\mathrm{bT}},
I_{\mathrm{b2}})$ at $I_{\mathrm{b1}}=-400$~$\mu$A. The coupling
strength between qubits 2 and 3 is $J_{23}/I_{p2}\Phi_0 =
M_{q2\mathrm{T}} \delta I_2^{*}/\Phi_0 \approx 0.034$.} \label{fig2}
\end{figure}

\begin{figure}[tbp]
\centering \includegraphics[scale=0.27]{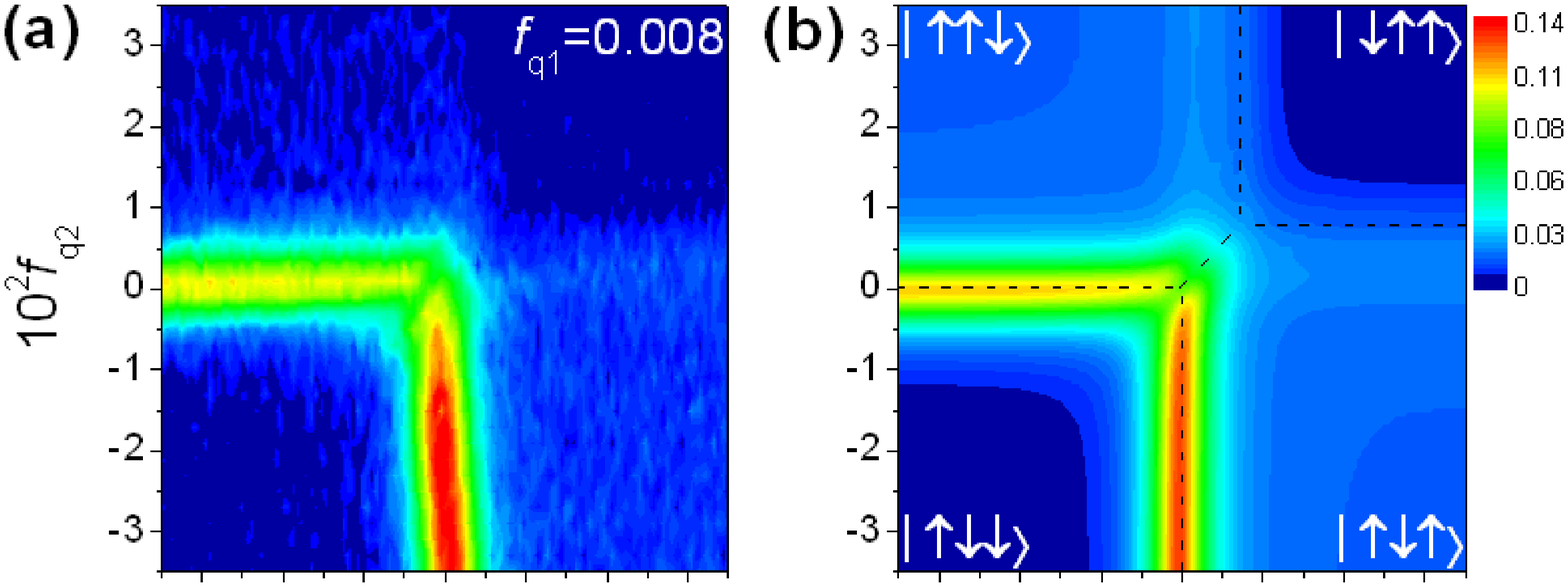} \centering
\vspace{1mm}
\includegraphics[scale=0.27]{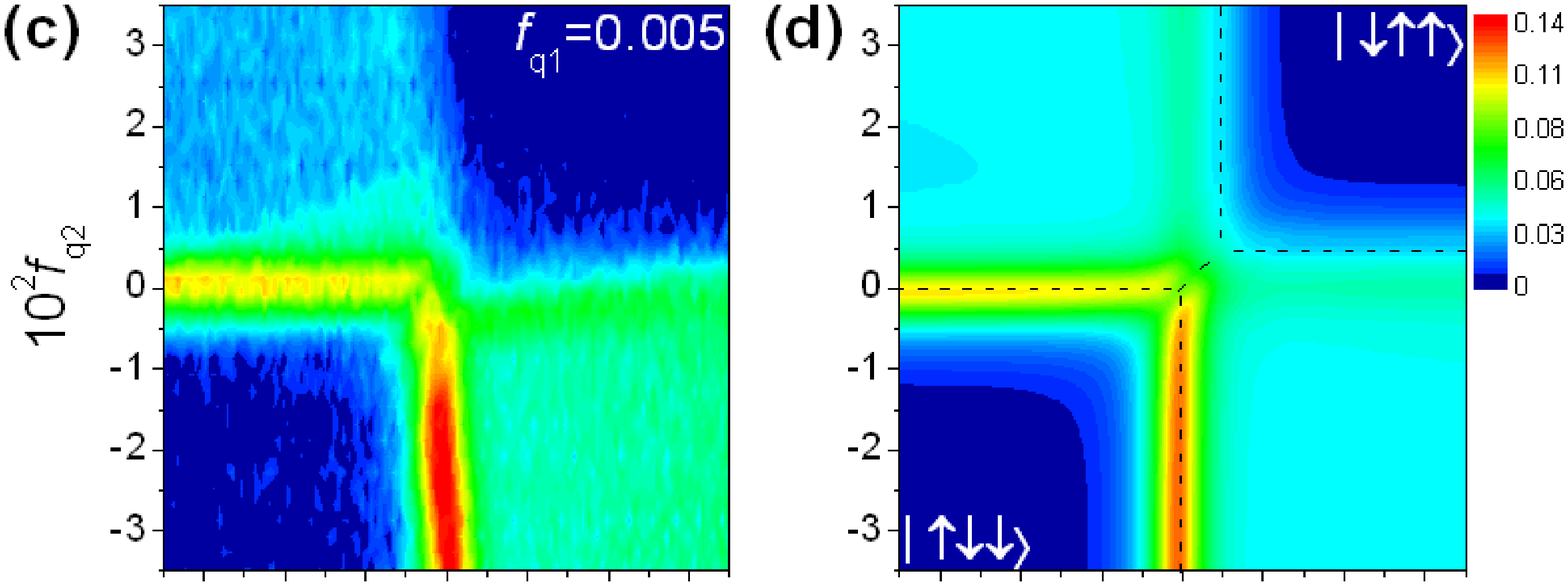} \vspace{1mm} \centering
\includegraphics[scale=0.27]{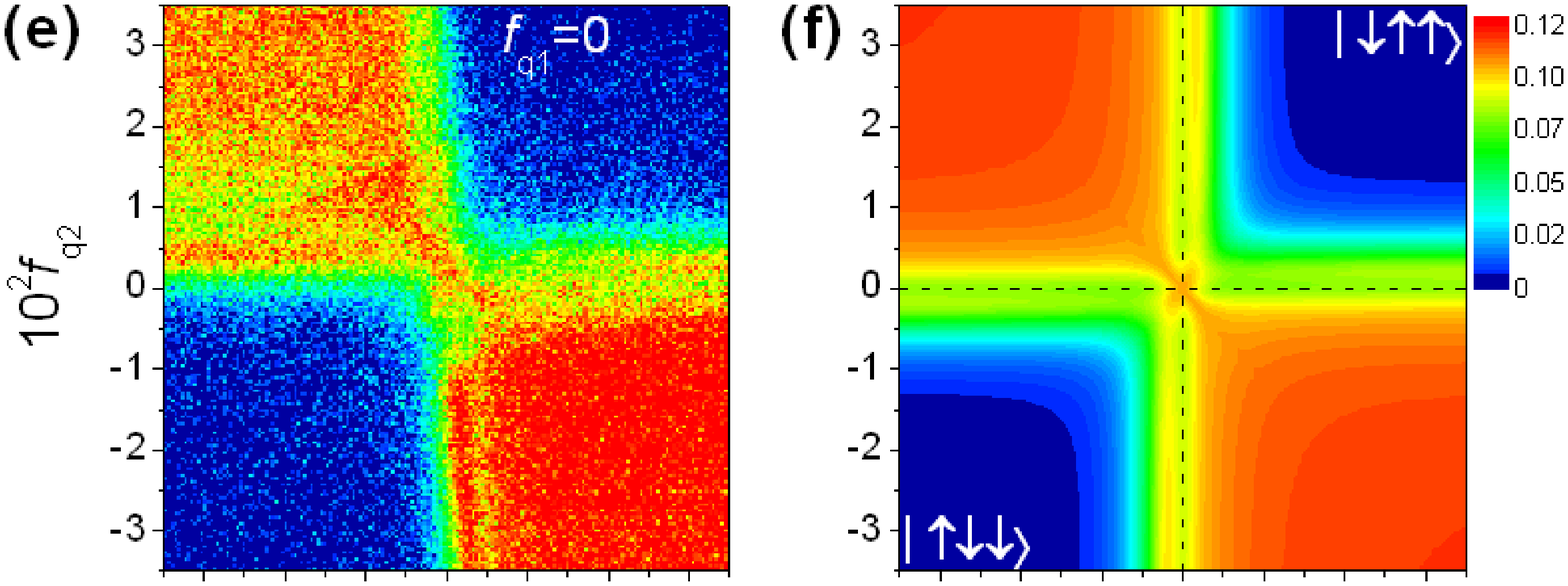} \vspace{1mm} \centering
\includegraphics[scale=0.27]{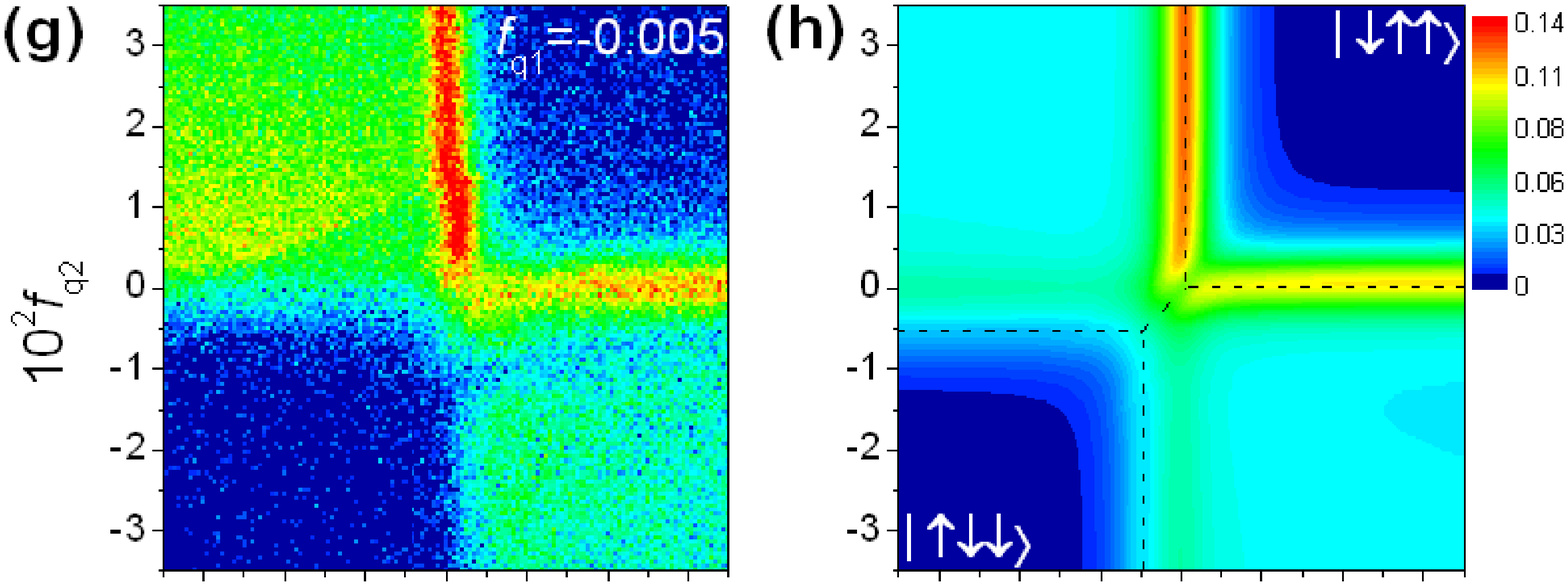} \vspace{1mm} \centering
\includegraphics[scale=0.27]{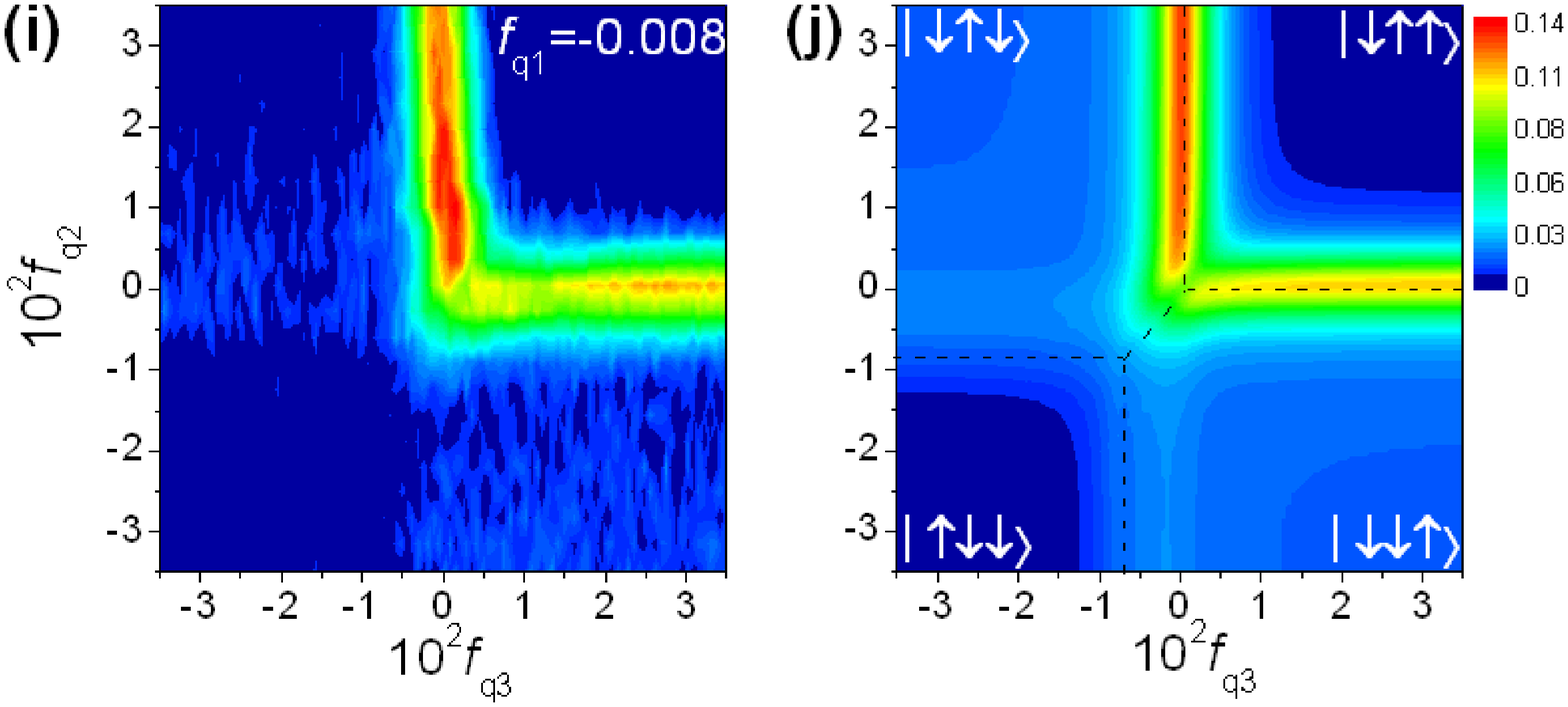}
\caption{(Colour on-line) Plot of $-\tan \theta (f_{q2}, f_{q3})$
for different fluxes $f_{q1}$ through qubit 1. The left column (a,
c, e, g, i) represents the tank circuit phase shift, measured for a
mixing chamber temperature of 10 mK, while the right column (b, d,
f, h, j) is the theoretical prediction for a qubit system at
$T_{\mathrm{eff}}=70$~mK. The black dashed lines denote the
cross-overs between the different classical states. The imperfect
compensation visible in the upper left corner is due to the
saturation of one of the current sources, therefore the points there
have a slightly larger $|f_{q1}|$.} \label{fig3}
\end{figure}

The qubit-coil mutual inductances were extracted from the $\Phi_0$
periodicity of the $ac$-susceptibility of the individual qubit as
$M_{q1\mathrm{T}}\approx 45.77$~pH and $M_{q2\mathrm{T}}\approx
46.62$~pH, $M_{q3\mathrm{T}} \approx 45.79$~pH. Figure~\ref{fig2}
plots the tangent of the phase shift $-\tan \theta (I_{\mathrm{bT}},
I_{\mathrm{bj}})$ ($j=1,~2$), as measured at the mixing chamber
temperature of 10 mK.  The Hamiltonian itself was completely
reconstructed from a number of scans away from the common degeneracy
point. Each trace corresponds to a single qubit anti-crossing, while
the repulsion between them confirms the \emph{anti-ferromagnetic}
coupling between all pairs of qubits~\cite{Grajcar2005a}. The
individual qubit parameters (persistent currents $I_{\mathrm{p}i}$
and tunnelling amplitudes $\Delta_i$) were found from the shape of
the peaks, $-\tan \theta(f_{\mathrm{bT}})$, when all other qubits
were biased away from their degeneracy~\cite{DeltaReconstruction}
(as an example we can choose the curve $-\tan
\theta(f_{\mathrm{bT}})$ for $I_{\mathrm{b1}}=600~\mu$A and
$I_{\mathrm{b2}}=400~\mu$A). The coupling amplitude $J_{ij}$ between
all pairs of qubits can be obtained by measuring the peak-to-peak
distance $\delta f_{i}=M_{qi\mathrm{T}} \delta I_i
/\Phi_0$~\cite{Grajcar2005a}, see fig.~\ref{fig2}. Thus, the
dimensionless coupling energy is equal to  $\delta f_i$, i.e.
$J_{ij}/I_{\mathrm{p}i} \Phi_0=\delta f_i$. The reconstruction of
the parameters resulted in $\Delta_1 \approx \Delta_2 \approx
\Delta_3  = 70$~mK, $I_{\mathrm{p}1} \approx I_{\mathrm{p}2} =
115$~nA, $I_{\mathrm{p}3} = 125$~nA, $J_{12}/I_{p2}\Phi_0 =
M_{q2\mathrm{T}} \delta I_2/\Phi_0 = 0.034$, $J_{13}/I_{p3}\Phi_0 =
M_{q3\mathrm{T}} \delta I_3/\Phi_0 = 0.033$ and $J_{23}/I_{p2}\Phi_0
= M_{q2\mathrm{T}}  \delta I_2^{*}/\Phi_0 = 0.034$, at the effective
temperature $T_\mathrm{eff}=70$~mK of our setup known from previous
experiments~\cite{Grajcar2005a, Grajcar2005b}. The discrepancy
between effective and mixing-chamber temperature is due to the noise
from external leads and amplifier. The substitution of the
persistent currents into the later three expressions gives $J_{12}
\approx J_{13} \approx J_{23} = (0.61 \pm 0.02)$~K. The slope of the
traces in fig.~\ref{fig2} gives the mutual inductances between
qubits and \textit{dc}-bias lines. We found $M_{q1,\mathrm{b1}}=
0.595$, $M_{q2,\mathrm{b1}}= 0.387$, $M_{q3,\mathrm{b1}}= 0.329$,
$M_{q1,\mathrm{b2}}= 0.296$ and $M_{q2,\mathrm{b2}}= 0.347$,
$M_{q3,\mathrm{b2}}= 0.543$ (all in pH).

The experimental reconstruction of the mutual inductance matrix
$M_{qi,\mathrm{bi}}~(\mathrm{bi}=\mathrm{b1},~\mathrm{b2},\mathrm{bT})$
facilitates the possibility of realising automated flux control for
each qubit. Indeed, if we want to set the fluxes $(f_{q1}^0,
f_{q2}^0, f_{q3}^0)$ in the qubits, we have to feed the bias lines
with the direct currents: $I_{\mathrm{bj}}^0=\Phi_0 \sum_i
M^{-1}_{qi,\mathrm{bj}} f_{qi}^0$. Figure~\ref{fig3} shows $-\tan
\theta (f_{q2}, f_{q3})$ for different fluxes through qubit 1. The
left column (a, c, e, g, i) represents the tank circuit phase shift,
measured at a base temperature of 10 mK, while the right column (b,
d, f, h, j) is the theoretical prediction~\cite{Izmalkov2004a} for
the sample parameters determined from the fig.~\ref{fig2}. The
experimental and theoretical data are in good agreement.

Figure~\ref{fig3} can be understood if one looks at the ground state
in the classical limit ($\Delta_1 \rightarrow 0,~\Delta_2
\rightarrow 0,\Delta_3 \rightarrow 0$). In that case we have sharp
crossovers between the basis states. In the figures they are
indicated by black dashed lines between the states
$\mathopen|\uparrow_1\uparrow_2\downarrow_3\rangle$,
$\mathopen|\downarrow\uparrow\uparrow\rangle$,
$\mathopen|\uparrow\downarrow\downarrow\rangle$ and
$\mathopen|\uparrow\downarrow\uparrow\rangle$ in fig.~\ref{fig3}b
and the states $\mathopen|\downarrow\uparrow\downarrow\rangle$,
$\mathopen|\downarrow\uparrow\uparrow\rangle$,
$\mathopen|\uparrow\downarrow\downarrow\rangle$ and
$\mathopen|\downarrow\downarrow\uparrow\rangle$ in fig.~\ref{fig3}j.
Here the spin-up and spin-down notation describes different
directions of the persistent current in the qubit loops. There is
also a crossover from $\mathopen|\uparrow\uparrow\downarrow\rangle$
and $\mathopen|\uparrow\downarrow\uparrow\rangle$  to
$\mathopen|\downarrow\uparrow\downarrow\rangle$ and
$\mathopen|\downarrow\downarrow\uparrow\rangle$ at the plane of
fig.~\ref{fig3}f ($f_{q1}=0$). The ferromagnetic states
$\mathopen|\uparrow\uparrow\uparrow\rangle$ and
$\mathopen|\downarrow\downarrow\downarrow\rangle$ are not reached
for this flux subspace, because of their relatively high energies.

\begin{figure}
\centering \includegraphics[width=5cm]{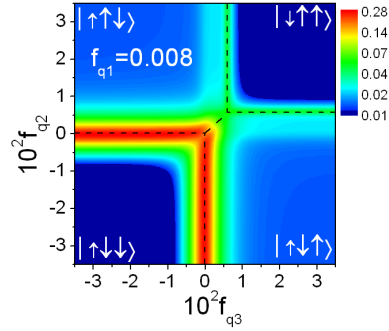} \caption{(Colour
on-line) $-\tan \theta (f_{q2}, f_{q3})$ at $f_{q1}=0.008$ for
hypotethical qubit parameters: $J_{12}=J_{13}=J_{23}=0.3$~K,
$I_{\mathrm{p}1}=130$~nA, $I_{\mathrm{p}2}=I_{\mathrm{p}3}=180$~nA,
$T_{\mathrm{eff}}=70$~mK. All other parameters are the same as in
experiment. Dashed lines denote the classical cross-overs between
the different two--qubit states.} \label{fig4}
\end{figure}

For finite $\Delta_i$ macroscopic quantum tunneling removes the
classical degeneracy and the qubit's wave function becomes a
superposition of the spin states in the vicinity of the
anti-crossing. As a result the ground and first excited state of the
multi-qubit system exhibit significant curvature. In accordance with
formula~(\ref{formula1}) the phase shift is maximal at the
anti-crossings and from the measurements one would be able to
reconstruct the classical anti-crossing curves. Nevertheless, this
is not the case for all parameter space because we measure the
susceptibility with respect to the total flux. For the transitions
from $\mathopen|\downarrow\uparrow\uparrow\rangle$ to
$\mathopen|\uparrow\uparrow\downarrow\rangle$ or
$\mathopen|\uparrow\downarrow\uparrow\rangle$ the total
``magnetization'' does not change, thus there is no susceptibility
change and no phase shift. Moreover, for our set of parameters, in
the vicinity of these transitions the difference between the ground
and first excited state is smaller than the system's effective
temperature. Therefore the $ac$-susceptibility is suppressed because
of the partial occupation the first excited state
\cite{Grajcar2005a}. Nevertheless, the transitions would be visible
if the persistent currents were different as shown in fig.
\ref{fig4}. In this case the magnetization of the states mentioned
above would be slightly different resulting in a change of the
susceptibility at their crossover.

The very good agreement between the measured susceptibility and its
prediction underlines the usefulness of the parametric transducer
for characterising qubits in the ground state. By using this
equilibrium measurement technique we can find some of the
anti-crossings for the flux space $\mid f_{q1} \mid \gtrsim 0.003$
and all for $\mid f_{q1} \mid \lesssim 0.003$. The range where
complete characterization is possible can be even increased by
either decreasing the effective temperature of the setup, or by
fabricating another sample with a higher $\Delta/J$ and qubits with
different critical currents. As an illustration we calculated the
system response (see fig.~\ref{fig4}) for hypothetical parameters
$J=0.3$~K, $I_{\mathrm{p}1}=130$~nA,
$I_{\mathrm{p}2}=I_{\mathrm{p}3}=180$~nA and with all the others the
same as for the current sample. Here all classical crossovers are
clearly visible even for $T_{\mathrm{eff}}=70$~mK. By using
non-equilibrium measurements with adiabatic evolution it should be
possible to determine the state completely. In this kind of
measurement one would prepare the system in the ground state at a
point where this state is easily reached and, subsequently, let it
evolve to the problem Hamiltonian adiabatically slowly but fast
enough to avoid thermal excitation \footnote{The upper limit for
adiabatic evolution/read-out speed $\lambda$ near minimal energy gap
$g_{min}$ point can be easily estimated from the Landau-Zener
formula~\cite{LZ}: $\lambda \lesssim \frac{g_{min}^2}{2 \hbar}$.}.

In conclusion, the three flux qubit susceptibility was completely
reconstructed. The experimental data are found to be in complete
agreement with quantum mechanical predictions in full parameter
space. We also demonstrate a \emph{controllable} multi-qubit
ground band anti-crossing read-out, which allows ground-state
computation with superconducting flux qubits. The next steps in
this direction would be the improvement of the read-out speed and
demonstration of adiabatic quantum computation and its efficiency.

\acknowledgments AI, SvdP, EI were supported by the EU through the
RSFQubit and EuroSQIP projects and MG by Grants VEGA 1/2011/05 and
APVT-51-016604. We thank E. Goldobin, R.~Gross, H.~E.~Hoenig,
S.~Linzen, M.~J.~Storcz, Th.~Wagner, and A.~Zeilinger for fruitfull
discussions.

\end{document}